# Observation and Simulation of All Angular Magnetoresistance Oscillation Effects in the Quasi-One Dimensional Organic Conductor (DMET)$_2$I$_3$

Pashupati Dhakal,[1] Harukazu Yoshino,[2] Jeong-Il Oh,[1] Koichi Kikuchi[3] and Michael J. Naughton[1]

[1] Department of Physics, Boston College, Chestnut Hill, MA 02467 USA

[2] Graduate School of Science, Osaka City University, Osaka 558-8585 Japan

[3] Graduate School of Science, Tokyo Metropolitan University, Tokyo 192-0397 Japan

Abstract

Measurements and calculations of magnetotransport in the molecular organic conductor (DMET)$_2$I$_3$ detect and simulate all known angular magnetoresistance oscillation (AMRO) phenomena for quasi-one dimensional (Q1D) systems. Employing the true triclinic crystal structure in the calculations, these results address the mystery of the putative vanishing of the primary AMRO phenomenon, the Lebed magic angle effect, for orientations in which it was expected to be strongest. They also show a common origin for Lebed and so-called "LN" oscillations, and confirm the generalized nature of AMRO in Q1D systems.

Quasi-one-dimensional (Q1D) molecular conductors are highly anisotropic materials which show remarkable oscillatory magnetotransport phenomena with respect to crystal orientation in a strong magnetic field [1,2]. Several types of angular magnetoresistance oscillations (AMRO) have been observed in Q1D conductors. In the prototypical Q1D conductors based on the TMTSF molecule, Lebed "magic angle" (LMA) resonances [3,4,5,6,7], Danner-Kang-Chaikin (DKC) oscillations [8, 9] and the Yoshino *et al.*-discovered angular effect (YAE) [10,11,12] have been observed for field rotations about the three principle axes $x//b$, $y//a'$ and $z//c^*$, respectively In addition, more complex "LN" oscillations [13] were observed when the magnetic field was rotated through arbitrary, out-of-plane directions [12,13,14].

While such AMRO effects have been detected in several Q1D materials, their origin(s) and relationships to each other have puzzled researchers for over two decades. For example, while numerical calculations of magnetoconductivity using the Boltzmann transport equation for a Q1D system qualitatively reproduce the observed AMRO [10], several other theoretical models introduced to explain interlayer AMRO in Q1D systems, quasi-classical and quantum [15,16,17,18,19,20,21,22,23,24], qualitatively explain only some of the observed effects (DKC, YAE, and LN). Curiously, these theories have consistently failed to simulate the initially predicted [3,4], and detected [5,6], Lebed effect. The models in Refs. 21-23 result in identical expressions for the interlayer conductivity, though from slightly different starting assumptions, each yielding a series of resistivity minima for integer values of an index *n* in the Lebed relation $\tan\theta = na/c$, where $\theta$ is the magnetic field angle between lattice directions *a* and *c*. According to these models, each $n^{th}$-order oscillation is modulated by an equivalent order Bessel function that is itself a function of the magnetic field ratio $B_x/B_z$, *x* and *z* being the intrachain and interplane (the most and least conducting) directions, respectively. However, when the field is rotated in the *y-z* plane (*i.e.* perpendicular to the Q1D chains, *x*), the presumed optimal situation for the Lebed effect, all Bessel functions vanish, with the exception of $n = 0$, and the resulting resistivity has a smooth, featureless angular variation with field, with *no* Lebed oscillations. In spite of this fact, the Lebed effect was recently been suggested to be the "only fundamental angular effect" [25], with all others (DKC, YAE, and LN) being modulations of it. This seems arguable since, experimentally, Lebed oscillation amplitudes have been anecdotally observed to *decrease* (some becoming immeasurably small) as the field rotation plane approaches the "preferred" *y-z* plane where, again, the effect is expected to be *strongest* [12,25].

To date, all available theoretical models [15-26] have employed an orthorhombic or cubic approximation to the actual triclinic crystal structure of the materials in which the AMRO effects have been seen. In this work, we have simulated conductivity via numerical calculations employing the *actual* triclinic lattice parameters of a Q1D conductor, $(DMET)_2I_3$, and measured its interlayer magnetoresistance. We show that all AMRO effects appear in both theory and experiment and, moreover, now match in the Lebed rotation plane with respect to the overall magnetoresistance and the presence of LMA features, including their still somewhat curious diminishment upon approaching the *y-z* plane.

$(DMET)_2I_3$ [27] shares many similarities with the TMTSF system, including Q1D AMRO effects, superconductivity, and a triclinic crystal structure. In this symmetry, the orthogonal set (*x, y, z*) is represented by (*b, a', $c^*$*), based on the lattice parameters *b, a* and *c* [11]. While the aforementioned AMRO effects have all been detected in $(TMTSF)_2X$, $(DMET)_2I_3$ may an be ideal material in which to study them, since it does not require high pressure to stabilize a metallic phase, and spin density waves appear much higher than in the TMTSF salts [28]. Also, $(DMET)_2I_3$ is the first organic conductor where the YAE was observed [11,29]. Interlayer resistance $R_{zz}(\theta,\phi)$ was measured on two crystals, each measuring $\sim 0.5 \times 0.3 \times 0.15$ mm$^3$, at 100 mK and 9 T using a dilution refrigerator and split-coil superconducting magnet equipped with a two axis ($\theta, \phi$) rotator. We show data for one sample, as both showed similar results. The polar angle $\theta$ was varied by rotating the magnet about the vertical while keeping the sample stationary, and an internal rotating stage controlled the azimuthal angle $\phi$ with a stepper motor-driven Kevlar string.

The measured angle-dependent magnetoresistance is shown in Fig. 1(a), as a function of $\theta$ for various values of $\phi$. Data were recorded every half degree in $\theta$, from $-100°$ to $100°$, and every five degrees in $\phi$, from $0°$ to $180°$. Large AMRO are seen throughout. Although not shown in detail here, clear DKC oscillations are observed (previously unreported for this material), when rotating near $|\theta| = 90°$ at $\phi = 0°$ and $180°$. LMA oscillations are also observed for $\theta$-rotations when $\phi = 90°$ (dotted line). The YAE is clearly observed (again not detailed) for a $\phi$ rotation with fixed $|\theta| = 90°$. Finally, the oscillations that appear for virtually all $\phi$ rotations with fixed $\theta$ are manifestations of the LN effect. Thus, *all* previously-reported AMRO effects are experimentally observed in the present single experiment, as indicated in the figure. Next, we compare the experimentally-observed results of Fig. 1(a) with new calculations based on the actual triclinic crystal structure, Fig. 1(b), as well as with the existing theoretical models that use an orthorhombic approximation, Fig. 2.

The magnetoconductivity tensor is calculated by solving the one-electron Boltzmann transport equation within the relaxation time approximation for the triclinic crystal structure of (DMET)$_2$I$_3$ [30]. The equation is given by

$$\sigma_{ij} = \frac{2e^2}{V} \sum_k \left(-\frac{df}{dE}\right) v_i(k,0) \int_{-\infty}^{0} v_j(k,t) e^{t/\tau} dt, \qquad (1)$$

where $e$ = electronic charge, $V$ = sample volume, $f$ = Fermi distribution function, $E$ = electron energy, $v_i$ = $i^{\text{th}}$-component of the carrier velocity, $k$ = electron wave vector, $t$ = time, and $\tau$ =

relaxation time, respectively, with $\tau$ assumed to be constant. The carrier velocity is calculated based on the tight binding energy dispersion, $E = -2t_b \cos k_b b - 2t_a \cos k_a a - 2t_c \cos k_c c$, where $t_b$, $t_a$ and $t_c$ are transfer integrals along the respective lattice directions.

We calculate the interlayer magnetoresistance using bandwidth ratios [31] $t_b : t_a : t_c = 300 : 30 : 1$ and $\tau = 10^{-14}$ s. The velocities $v_b$, $v_a$ and $v_c$ are calculated and transformed along the $x$, $y$ and $z$-axes of the Cartesian coordinate system by using matrix transformations. Once the Lorentz force and wave vectors are calculated in the Cartesian coordinate system, they are converted to the triclinic system by using an inverse matrix transformation. This calculates the new Fermi velocity from the dispersion relation. To acquire results with sufficient precision, the first Brillouin zone is divided into a grid of 128 x 128 x 128 sites. Figure 1(b) shows the resulting calculated interlayer magnetoresistivity ($\rho_c = \rho_{zz} \cong 1/\sigma_{zz}$), again as a function of $\theta$ at the same angles $\phi$ as for the experiment. Here, it can be seen that the calculated magnetoresistance is qualitatively and semi-quantitatively in accordance with the experimental data, reproducing all known AMRO effects. The angular positions of magnetoresistance minima for various integer indices $n$ can be defined by what can be viewed as a generalized Lebed+LN relation, which we'll term LNL:

$$\tan\theta \sin\phi = n\frac{a\sin\gamma}{c\sin\alpha \,\sin\beta^*} + \cot\beta^* \qquad (2)$$

where the integer $n$ is the Lebed (or LNL) index, and $\cos\beta^* = (\cos\gamma \cos\alpha - \cos\beta)/(\sin\alpha \sin\gamma)$, using the crystal lattice angles $\alpha$, $\beta$ and $\gamma$.

Figure 2 shows magnetoresistance calculations for $(DMET)_2I_3$ in the Lebed $y$-$z$ rotation plane ($\phi = 90°$). Figure 2(a) uses orthorhombic crystal symmetry (for both numerical Boltzmann and analytical Kubo formalisms [21]), while Fig. 2(b) uses the present triclinic symmetry

Boltzmann model. These can be compared to our experimental data in Fig. 2(c). The positions of Lebed minima calculated from Eq. 2, using the triclinic lattice parameters for $(DMET)_2I_3$ [32], are in good agreement with the experimental results of Fig 2(c). Moreover, in contrast to previous Kubo-based theoretical models (smooth solid curve with no oscillating features in Fig. 2(a)), we show that the Lebed effect appears in Boltzmann-type calculations, regardless of crystal symmetry (dashed curves with oscillating features in Figs. 2(b) and (c)). In particular, our new triclinic Boltzmann calculations, Fig. 2(b), reveal Lebed effect features on the scale shown for indices as high as $n = 5$, with oscillation amplitudes similar to the experimental results and more than ten times larger than the orthorhombic calculations of Fig. 1(a). Derivatives $\partial^2\rho/\partial\theta^2$ illustrate this latter point, as well as the complete absence of LMA in the Kubo model calculations [21].

There is broad agreement as well between the calculated positions of resistance minima for the generalized LNL effect, given by Eq. 2, and experimental data, as shown in Fig. 3. Here, the LNL minimum index $n$ is plotted as a function of $\tan\theta$ for in-plane ($\phi = 90°$, LMA) and out-of-plane ($\phi \neq 90°$, LN) rotations. A series of symmetric patterns emerges, not about $\tan\theta = 0$ but about $\tan\theta \approx -0.15$. Moreover, these minimum positions shift progressively away from this symmetry point as rotations move away from the $y$-$z$ plane (*i.e.* as $\phi$ deviates from 90°) as indicated by the curved lines. Meanwhile, the calculated index curves $n[\tan\theta]$ (straight, solid lines) are symmetric about $\tan\theta_{n=0} = -\cot\beta^* = -0.146$, owing to the triclinicity of the crystal structure (Eq. 2). This agrees with the above −0.15 experimental value. Figure 3 provides compelling evidence that the LMA and LN effects are in fact two aspects of the same effect, with the same underlying physics, providing rationale for the LNL moniker.

From the present calculations and experiments, it is seen that all known Q1D AMRO effects are indeed observed in $(DMET)_2I_3$, and by extension, are generic to Q1D systems with like crystal and band structures. The *x-z* plane DKC effect is connected directly to the details of the warped Fermi surface through the transfer integral ratio $t_y/t_x$ [8], whereas the *x-y* plane YAE is ascribed to the velocity-preserving nature of "effective" electrons, via their proximity to the geometrical inflection points on the Fermi surface [13,33], which can be used to estimate the interplane coupling $t_z$. The *y-z* plane LMA effect, which prompted initial investigations of Q1D AMRO, can be explained in terms of the commensurability of electron trajectories across the warped Fermi surface, but the positions of the LMA magnetoresistance minima are determined only by the lattice parameters (Eq. 2). The more general form of LMA is the LNL effect. There appear to be, therefore, three (formerly four) distinct Q1D AMRO angular effects. The calculations do not reproduce every aspect of the measured resistivity, however. In particular, it can be seen in Figs. 1 and 2 that the measured data exhibit a broad minimum at $B//y$ ($|\theta|=90°$), while calculations show a local maximum expected from Lorentz force considerations. While tempting to associate this behavior with anomalous superconductivity for this orientation, as is seen in $(TMTSF)_2PF_6$ [34], further studies will be required to address this issue.

Detailed analyses of the results of Fig. 1 reveal that the amplitudes of most LNL oscillations indeed diminish rapidly as the rotation plane approaches the LMA orientation ($\phi=90°$). Triclinic Boltzmann numerical calculations show diminishing (but finite) LNL oscillation amplitudes for all indices *n* while approaching the *y-z* plane, as shown in Fig. 4(b), while for the Kubo analytic model, all oscillations completely vanish within our calculation precision. We have also determined the amplitudes of the experimentally-observed LNL features, with the results shown in Fig. 4(a). The curves in Fig. 1(a) are fit to a smooth background function ($\propto \sin^2\theta$), deviations from which are taken as oscillation amplitudes. While approximate, this procedure captures the overall dependence of the AMRO oscillations. As seen, the measured amplitudes indeed decrease, at least for $|n|>1$, while approaching the *y-z* plane, where naively they were expected to be maximal. One possible explanation for this behavior, now observed in both theory and experiment, is that as the magnetic field is tilted away from the Lebed plane, Fermi surface electrons acquire velocity components along the magnetic field which are even larger that those thought to be responsible for the original Lebed effect, resulting in stronger conductivity increases (deeper resistivity minima) for the generalized LNL effect. Likewise, detailed differences between theory and experiment may suggest that the one-electron theory employed may be too simplistic, and electron interactions, which are expected to increase in such low dimensional systems, may be involved.

In summary, we have measured the interlayer magnetoresistance of the Q1D organic molecular conductor $(DMET)_2I_3$ at low temperature and across all magnetic field orientations. All known Q1D AMRO effects are now observed in this system. We have numerically solved the interlayer magnetoconductivity tensor for the same field orientations, using the true triclinic lattice parameters, a procedure that should now be employable for other Q1D systems. Even though the LNL amplitudes decrease while the magnetic field rotation plane approaches the *y-z* plane, the calculated results confirm that Lebed oscillations survive, up to at least $n=5^{th}$ order, in contrast to some previous theoretical models which predict their absence. These Lebed oscillations may indeed be "magic" in Q1D molecular conductors.

This work was support by the National Science Foundation, under Grant No. DMR-0605339.

Figure Captions

FIG. 1. (color online) (a) Interlayer magnetoresistance of $(DMET)_2I_3$ versus polar angle $\theta$ for different azimuthal angles $\phi$ at $B = 9$ T and $T = 100$ mK. (b) Calculated magnetoresistance at 9 T using the triclinic crystal structure. Inset shows definitions of axes and field tilt angles. All known types of AMRO oscillations, LMA (dotted line at $\phi = 90°$), DKC, YAE and LN, as indicated, are detected in the experiment and reproduced in the calculations (DKC and YAE are clearly evident on expanded scales). The DKC effect had not been previously reported in this compound.

FIG. 2. (color online) Polar angle dependence (*i.e.* magnetic field rotated in *y-z* plane) of the 9 T magnetoresistance: (a) Orthorhombic Boltzmann numerical (solid) and Kubo analytic (dashed) calculations [21]; (b) Present triclinic Boltzmann numerical calculation; (c) Experiment. Insets show $d^2\rho/d\theta^2$ for the calculations, indicating the lack of Lebed oscillations using the Kubo formula, and their presence in Boltzmann calculations, with triclinic symmetry yielding features ~10 times larger than orthorhombic. Oscillations up to $n = 5$ or higher are observed in triclinic calculations (b) and experiment (c). Triangles indicate angular positions of the resistivity minima for indices $n$ according to Eq. 2.

FIG. 3. (color online) Positions of experimental magnetoresistance minima (circles), indexed by $n$ as in Fig. 2, versus $\tan\theta$ for different $\phi$ rotations. Solid lines are $n[\tan\theta]$ from Eq. 2 (thick line indicates Lebed plane, $\phi = 90°$), showing agreement between this generalized relation and the observed LNL minima, including the angular offset $\cot\beta^*$ due to triclinicity. Data and lines for different $\phi$ are offset for clarity.

FIG. 4. (color online) Amplitudes of magnetoresistance oscillations from data and simulations in Fig. 1 for different LNL indices $n$ versus angle $\phi$ as observed experimentally (a) and from present triclinic calculations (b). High index amplitudes decrease significantly while approaching the magic angle orientation $\phi = 90°$, but they remain finite.

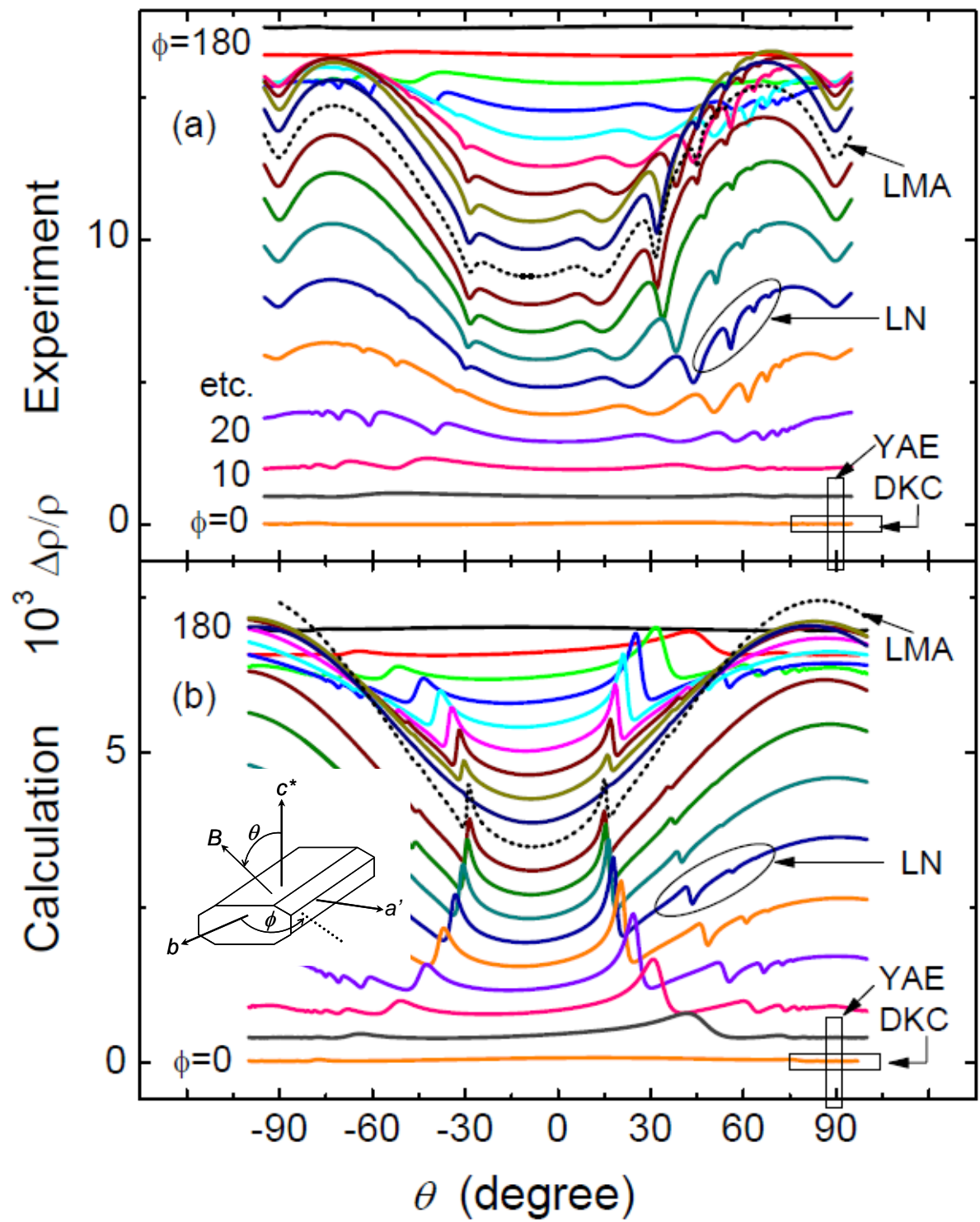

FIG. 1

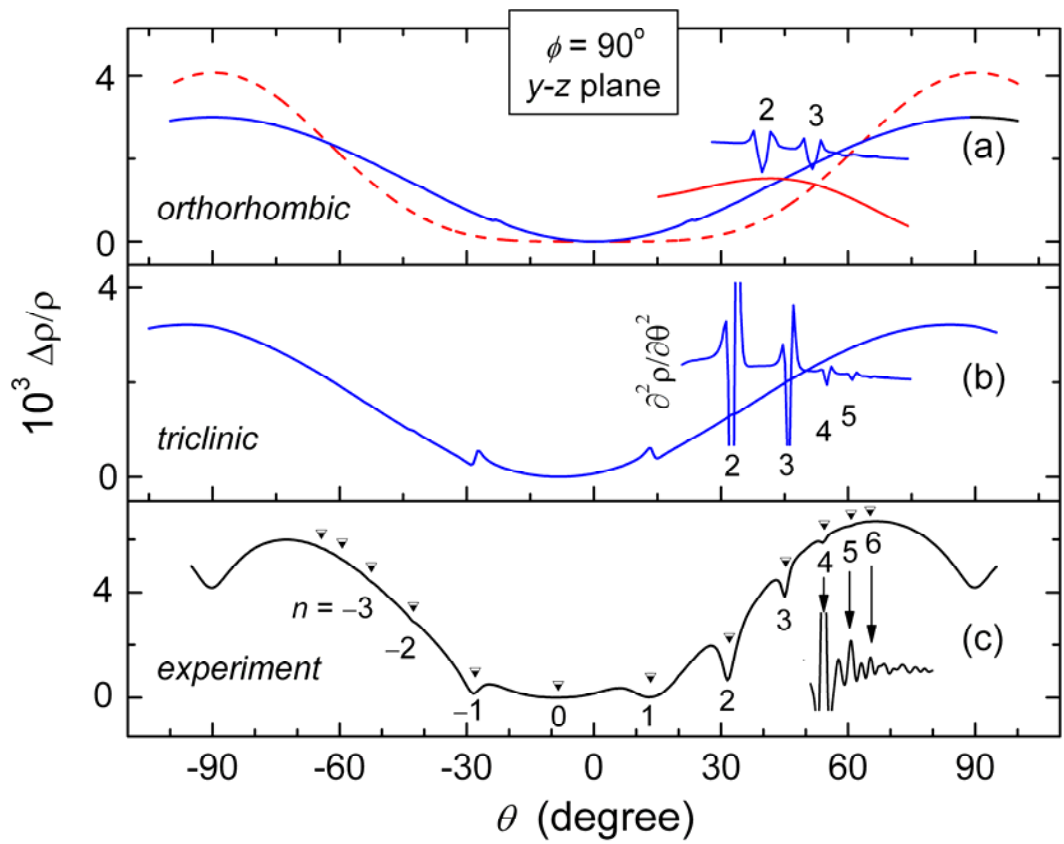

FIG. 2

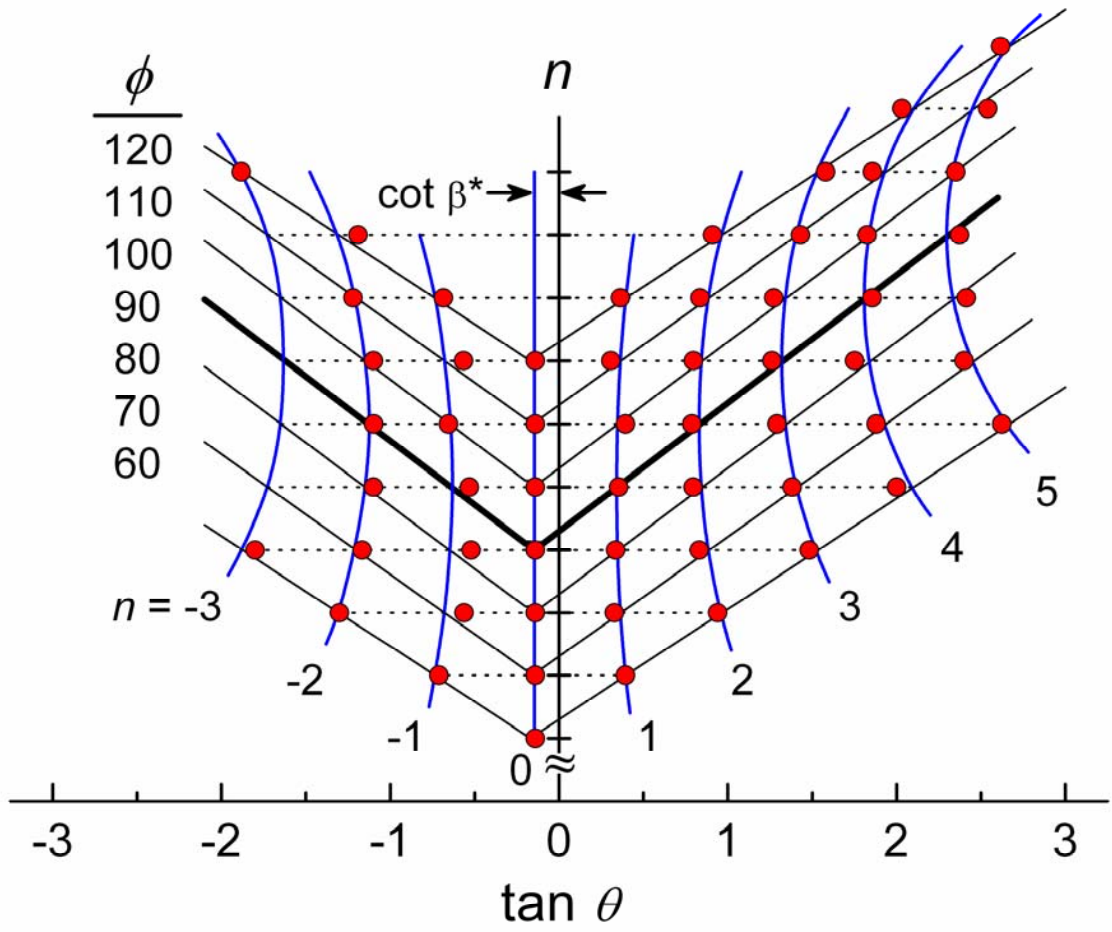

FIG. 3

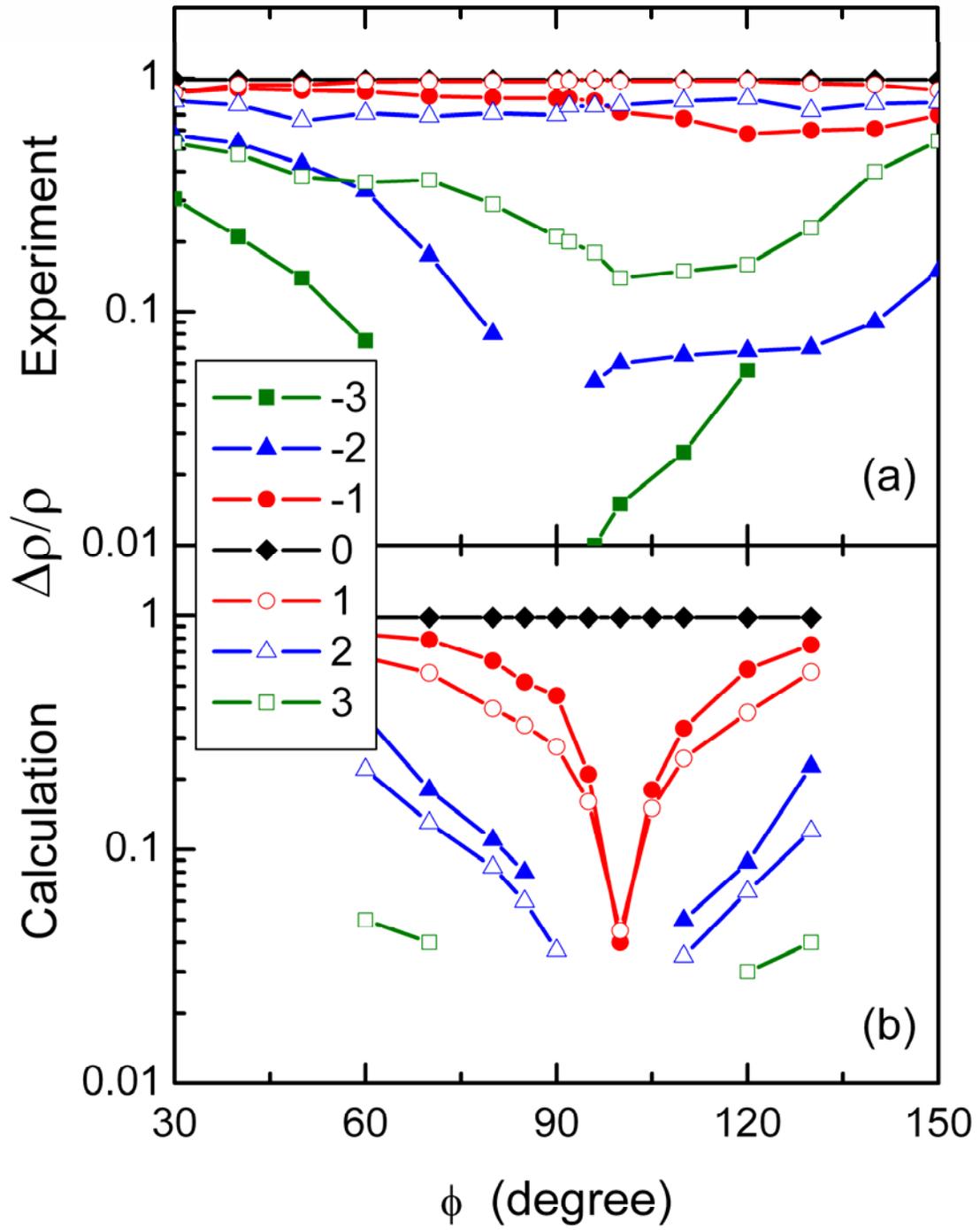

FIG. 4